\documentclass[]{spie}  

\usepackage{amsmath,amsfonts,amssymb}
\usepackage{graphicx}
\usepackage[colorlinks=true, allcolors=blue]{hyperref}

\providecommand{\arcsec}{\ensuremath{^{\prime\prime}}}

\title{MATISSE-py: a high-level Python interface for efficient, automated VLTI/MATISSE data reduction}

\author[a]{A. Soulain}
\author[a]{A. Matter}
\author[a]{P. Berio}
\author[a]{F. Millour}
\author[a]{B. Lopez}
\author[a]{R. Petrov}
\author[a]{S. Lagarde}
\affil[a]{Universit\'e C\^ote d'Azur, Observatoire de la C\^ote d'Azur, CNRS, Laboratoire Lagrange, Bd de l'Observatoire, CS 34229, 06304 Nice cedex 4, France}

\authorinfo{Further author information: (Send correspondence to A.S.)\\A.S.: E-mail: anthony.soulain@oca.eu}

\begin{document}
\maketitle

\begin{abstract}
The Multi AperTure mid-Infrared SpectroScopic Experiment (MATISSE) at the Very Large Telescope Interferometer (VLTI) delivers rich mid-infrared interferometric datasets that are reduced through a robust, multi-stage ESO pipeline. In its native form, however, the reduction relies on command-line calls to \texttt{esorex}, hand-edited set-of-files (SOF) and configuration files, and manual iterations that are difficult to automate, reproduce, or integrate into modern Python workflows. We present MATISSE-py, a modular Python interface that wraps the official ESO \texttt{esorex} recipes behind a single, user-friendly \texttt{matisse} command. It automates discovery, reduction, calibration, and inspection of the data, and ships as a tested, open-source package built on industry-standard practices. Beyond convenience, MATISSE-py introduces science-grade improvements developed over seven years of MATISSE operations: visibility- and photometry-factor (V-factor and P-factor) corrections derived from the GRAVITY fringe tracker to mitigate residual visibility losses, per-baseline data filtering, beam-commuting-device (BCD) ``magic-number'' corrections, and an absolute spectrophotometric flux calibration using state-of-the-art reference spectra. On faint targets observed under degraded conditions, the V-factor correction recovers up to a few percent of lost visibility and reduces the statistical uncertainties by nearly 20\%, while yielding calibrator diameters that are consistent across BCD modes. MATISSE-py thus consolidates the consortium's expertise into a maintainable package and shortens the reduction-to-science loop for the community.
\end{abstract}


\section{INTRODUCTION}
\label{sec:intro}

MATISSE, the mid-infrared spectro-interferometer of the VLTI, combines four telescopes to deliver spectrally dispersed visibilities, closure phases, correlated fluxes, and spectra across the L, M, and N bands.\cite{Lopez2022} These observables drive a broad science program, from the inner regions of protoplanetary disks to the dusty environments of active galactic nuclei and evolved stars. The richness of the data products---multiple BCD configurations, chopping modes, correlated fluxes, and multi-wavelength coverage---makes MATISSE a uniquely capable but also a complex instrument to work with.

Raw MATISSE data are processed by the official ESO Data Reduction Software, executed through the \texttt{esorex} recipe-running tool.\cite{Esorex} While robust and well validated, this native workflow relies on command-line invocations, manually prepared set-of-files and configuration files, and repeated iterations over the many available modes. This approach is hard to automate over a full observing night, difficult to reproduce exactly, and awkward to integrate into the Python-based analysis environments that most astronomers now use. The result is a steep learning curve that effectively restricts efficient data reduction to a handful of expert users.

To address this, we developed MATISSE-py,\cite{Soulain2026} a modern, modular Python interface that wraps the official \texttt{esorex} recipes behind a single, user-friendly \texttt{matisse} command. This work did not start from scratch: it builds on the various Python scripts developed by the consortium over the years since MATISSE first light in 2019. MATISSE users will recognize the \textit{mat-tools} set of functions used so far to reduce and calibrate the data. MATISSE-py consolidates this legacy and turns it into a maintainable, industry-standard package: it automates discovery, reduction, calibration, and inspection, and ships as a tested, open-source tool distributed on PyPI (version 0.8.0 at the time of writing, supporting Python 3.10--3.14 under an MIT license). Beyond safeguarding the consortium's operational knowledge, the goal is to expose science-grade calibration improvements to non-expert users. This paper describes the software architecture (Sec.~\ref{sec:arch}), the reduction workflow (Sec.~\ref{sec:workflow}), the enhanced calibration corrections (Sec.~\ref{sec:calib}), and a first validation of their scientific benefit (Sec.~\ref{sec:results}).

\section{SOFTWARE ARCHITECTURE}
\label{sec:arch}

\begin{figure} [ht]
   \begin{center}
   \includegraphics[width=0.95\textwidth]{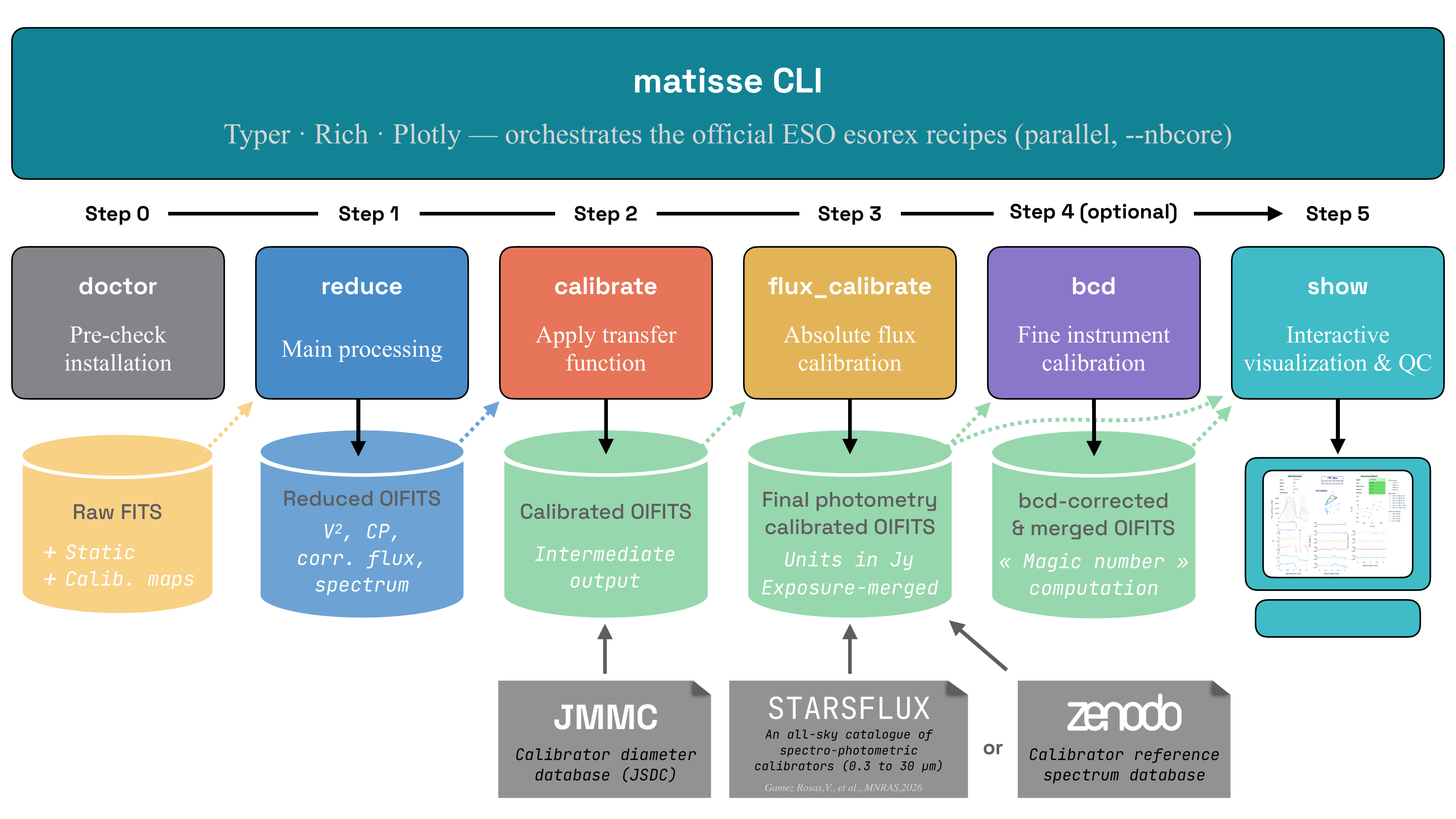}
   \end{center}
   \caption[workflow]
   { \label{fig:workflow}
The MATISSE-py architecture and data-reduction workflow. A single \texttt{matisse} command (built with Typer and Rich) orchestrates the official ESO \texttt{esorex} recipes, executed in parallel. The modular subcommands (colored squares) form a pipeline, each consuming and producing well-defined data products (light tube): raw FITS frames are reduced to OIFITS, calibrated against a matching calibrator, BCD-corrected and merged, flux-calibrated, and finally inspected. External reference databases (grey) feed the calibration steps, while \texttt{doctor} provides a pre-flight check of the environment.}
   \end{figure}

MATISSE-py follows a single-entry-point design: one \texttt{matisse} command exposes a set of modular subcommands, each mapping onto a stage of the reduction. The command-line interface is built with Typer\cite{Typer} and Rich for structured, self-documenting commands and informative terminal output, while interactive inspection relies on Plotly\cite{plotly}. Underneath, the package orchestrates the official \texttt{esorex} recipes, so that the numerical core of the reduction remains identical to the validated ESO pipeline. Figure~\ref{fig:workflow} summarizes this architecture.

A key design constraint is backward compatibility. MATISSE-py automatically detects the installed \texttt{esorex} recipe version and maps its high-level flags onto the appropriate low-level recipe parameters. New science corrections are exposed cleanly through descriptive options, with graceful fallback when a given recipe version does not support a feature. This keeps the official recipes and legacy reduction scripts usable while letting users adopt the high-level interface incrementally.

The package is engineered for reliability and long-term maintenance, following practices that are standard in software engineering but still uncommon in instrument pipelines: continuous integration and deployment, static type checking with mypy, linting with Ruff, a unit-test suite with coverage reporting (Pytest, Codecov), conventional commits and automated versioning (Commitizen), and fully automated deployment. This investment is deliberate: it converts knowledge that previously lived in individual scripts and expert practice into a tested, documented, and reproducible code base.

\section{THE REDUCTION WORKFLOW}
\label{sec:workflow}

The reduction is organized as a sequence of subcommands that take the user from raw frames to calibrated, inspectable data products.

\textbf{\texttt{matisse doctor}} checks the environment before any processing: it verifies the \texttt{esorex} installation, the available MATISSE recipes, and the calibration databases. This single diagnostic step removes a frequent and time-consuming source of setup errors.

\textbf{\texttt{matisse reduce}} converts raw FITS frames into reduced OIFITS files, extracting visibilities, closure phases, correlated fluxes, and spectra. The reduction steps run on \texttt{esorex} in parallel across multiple cores (\texttt{-{}-nbcore}), and the output is automatically formatted, so that an entire night can be reduced with a single command. Intermediate data products are organized into observing blocks, each corresponding to an individual MATISSE observing sequence (one BCD cycle, N-band photometry with chopping). Every block therefore contains a number of intermediate FITS files that remain available for detailed inspection. In the standard workflow, however, the user simply inspects the OIFITS files gathered in a single \texttt{reduced/} directory, named according to the MATISSE convention that encodes the BCD position, band, chopping status, and observation time.

\textbf{\texttt{matisse calibrate}} matches each science target to its calibrator in time and applies the corresponding transfer function to remove atmospheric and instrumental effects. The calibrated data products are science-ready from an interferometric point of view, but are still split into multiple exposures (i.e., the individual BCD positions). An additional step can therefore be performed to merge the different BCD positions into a single final OIFITS file (step 4, Fig.~\ref{fig:workflow}), and the correlated and total fluxes can be calibrated with \texttt{matisse flux\_calibrate}.

\textbf{\texttt{matisse flux\_calibrate}} performs the absolute flux calibration of the correlated  and total spectra. This requires the knowledge of the measured or modeled spectra of spectrophotometric calibrators. Several spectra databases are queried by \texttt{matisse flux\_calibrate} in the following order: 
\begin{itemize}
    \item the STellar Absolute Reference Spectroscopic Flux Library (STARSFLUX), an all-sky catalogue of optical-infrared spectra comprising almost 65,000 stars (Gamez et al., submitted)\cite{2026arXiv260711730G}. The list of calibrators is based on the Mid-infrared stellar Diameters and Fluxes compilation Catalogue (MDFC, VizieR II/361)\cite{Cruzalebes2019}, while the spectra are based on the NewEra PHOENIX models \cite{Hauschildt2025}. 
    \item a catalogue of brighter calibrators built by Jozsef
    Varga (private communication), and which contains 1665 stars taken from the MDFC. Here the spectra are based on the HiResFITS grid of PHOENIX stellar atmosphere models\cite{Husser2013}.
    \item finally a database described by van Boekel\cite{2004PhDT........79V}, which contains 482 objects, mostly Cohen calibrator \cite{Cohen1999}, and was originally used to calibrate VLTI/MIDI data. 
\end{itemize}

Note that any airmass and humidity level (pwv) difference between the science target and calibrator are corrected by dividing each raw flux by a synthetic atmospheric transmission profile computed with the Skycalc \cite{Noll2012, Jones2013}.

\textbf{\texttt{matisse bcd}} handles the beam-commuting-device modes through four operations: \texttt{compute} derives the relative instrumental contrast loss factor between the different fringe peaks and BCD configurations (the so-called magic numbers); \texttt{apply} simply applies them to every fringe peak of every BCD configuration to remove the instrumental contribution; \texttt{remove} re-orders the OIFITS tables to match the reference BCD ordering (OUT--OUT, beam not switched); and \texttt{merge} combines the different BCD positions into a single averaged OIFITS file.

\textbf{\texttt{matisse show}} provides an interactive Plotly viewer of the reduced OIFITS dataset, displaying target metadata, $uv$ coverage, spectra, visibilities, and closure phases, together with quality-check diagnostics. Inspection is therefore part of the same workflow rather than a separate, ad-hoc step.

\section{ENHANCED CALIBRATION}
\label{sec:calib}

Beyond automation, MATISSE-py incorporates calibration refinements developed from seven years of MATISSE operations. These act at two levels: inside the \texttt{esorex} recipes and in dedicated post-processing.

At the recipe level (\texttt{esorex} 2.4 and later), the main addition is the correction of two distinct sources of visibility loss using quantities recorded by the GRAVITY fringe tracker that operates alongside MATISSE in the GRA4MAT configuration.\cite{Woillez2024} The \emph{visibility factor} (V-factor) corrects the contrast loss caused by the residual optical-path (piston) jitter that remains after fringe tracking: over a MATISSE integration this residual phase jitter partially decorrelates the fringes and lowers the measured visibility, an effect that is strongest in the L and M bands. The \emph{photometry factor} (P-factor) corrects the photometric normalization of the visibilities using the per-telescope fluxes measured by the fringe tracker, accounting for injection and flux-ratio variations between the beams that would otherwise bias the calibrated visibility. Both corrections are obtained from the complex coherence and photometric quantities derived by the GRAVITY(+) fringe tracker---recorded together with the MATISSE data---and interpolated onto the MATISSE observables.\cite{Gravity2017} The same framework supports per-baseline data filtering (\texttt{-{}-filter-mode}, \texttt{-{}-filter-baseline}), fringe-jump rejection, and improved spectral calibration, including a better resampling of the \texttt{skycalc} atmospheric model.\cite{Noll2012}.

In post-processing, the package can optionally apply the BCD magic-number correction to the uncalibrated OIFITS files, removing residual instrumental effects. This new advanced tool allows us to debias the per-baseline fringe peaks, whose position on the detector affects the MATISSE beam input. As the BCD was built precisely to improve the differential-phase and closure-phase signals, the \texttt{esorex} pipeline is designed around this specific instrumental feature. The calibration step therefore applies the transfer function per exposure -- that is, per BCD position -- which degrades the potential SNR gain that would result from combining several exposures of the same calibrator star. To overcome this software limitation, we propose an optional post-processing step, via the \texttt{matisse bcd} command, that computes the instrumental bias affecting the fringe peaks -- the magic numbers -- and subtracts them from the individual BCD positions. The user can then merge the different exposures into a single calibrator OIFITS file, thereby improving the SNR, and perform the transfer-function calibration of each BCD-removed file against a single reference. As the MATISSE pipeline is not designed for this, we recommend using an additional Python software called \texttt{genoca}\footnote{\url{https://github.com/pberio/GenOCA}} to perform the transfer-function correction. Originally developed for the SPICA instrument at CHARA, it is now compatible with various interferometric instruments worldwide. This new MATISSE reduction capability is currently under review and should be released in the coming months, together with the corresponding workflow tutorial on the matisse-py GitHub repository. Finally, we implement a state-of-the-art spectrophotometric flux calibration described in Sec.~\ref{sec:workflow}. A curated database of reference calibrators is made publicly available to ensure reproducibility. Together, these corrections target the dominant systematic errors in MATISSE data while remaining optional and traceable, so that the user always knows which corrections have been applied.

\section{VALIDATION AND RESULTS}
\label{sec:results}

We validated the new corrections on two calibrator stars, HD\,145250 and HD\,139329, observed the 2$^{\mathrm{nd}}$ of July 2022 in the same sequence and calibrated against each other. The two calibrators were observed within $\sim$33 min of each other under stable conditions, with a mean seeing of 0.86'' and a mean coherence time of 3.4 ms.

Their angular diameters are independently predicted from surface-brightness/color relations and tabulated in the JSDC catalogue.\cite{Bourges2017} The reduction was run with the full filtering enabled---V-factor, P-factor, and fringe-jump rejection. We then performed an internal consistency test that consisted in fitting the uniform-disk diameter of each calibrator separately for each BCD mode: a well-calibrated dataset should return a diameter that is stable across BCD modes and baselines, and compatible with the catalogue value.

\begin{figure} [ht]
\begin{center}
\includegraphics[width=0.65\textwidth]{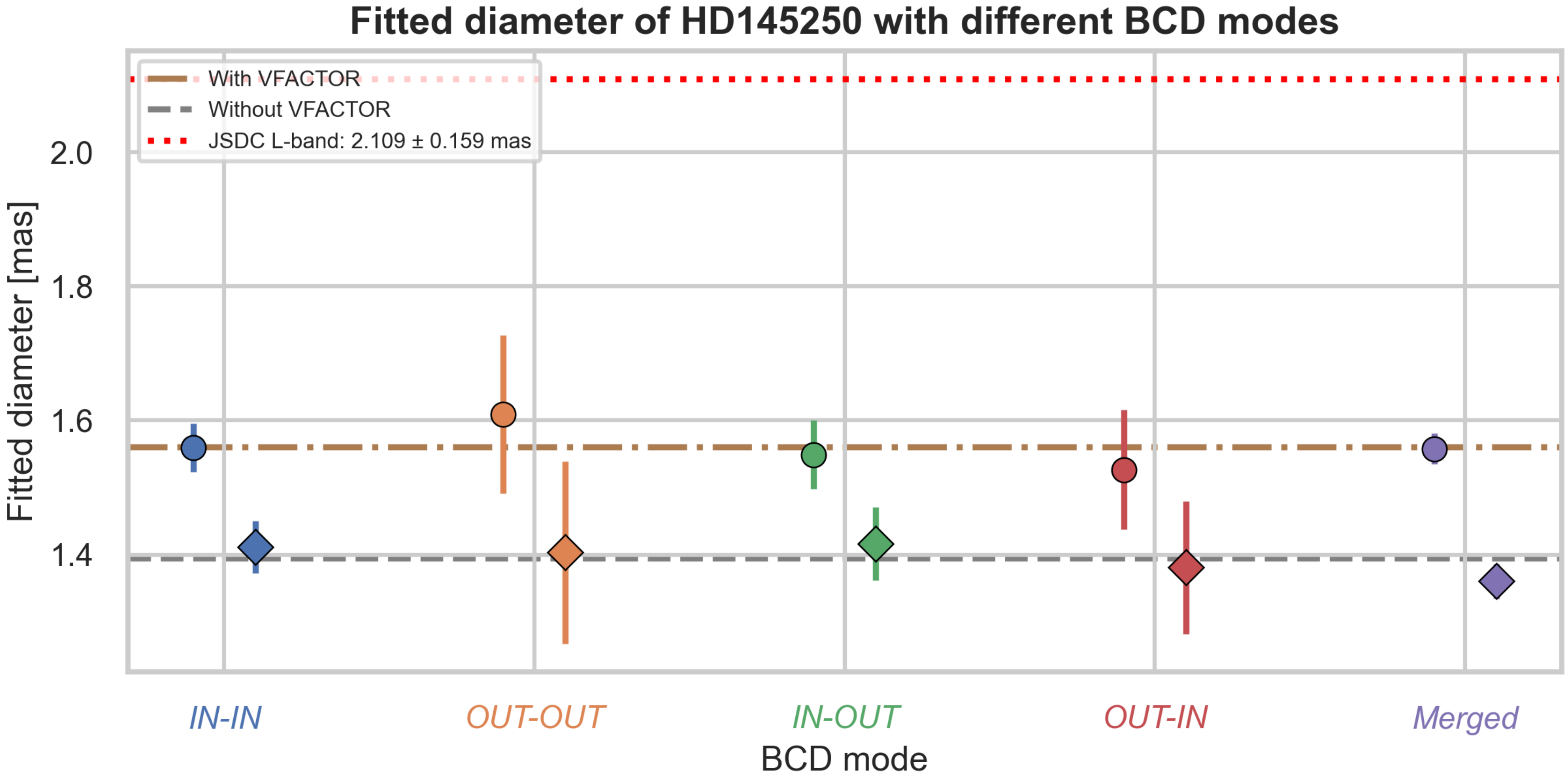}
\end{center}
\caption[diameter]
{ \label{fig:diameter}
    Uniform-disk diameter of the calibrator HD\,145250 fitted independently for each BCD mode in the L band, with (circles) and without (diamonds) the V-factor correction. Within the error bars the diameters are consistent across BCD modes in both cases; the main effect of the correction is to shift the fitted diameter upwards (brown vs.\ grey dash-dotted lines) towards the JSDC catalogue value\cite{Bourges2017} (dotted line), without fully reaching it.
}
\end{figure}

Figure~\ref{fig:diameter} shows this test for HD\,145250 in the L band. Within their uncertainties, the per-mode diameters are mutually consistent both with and without the correction, so the dominant effect of the V-factor is not a reduction of the BCD-to-BCD scatter but an overall shift of the fitted diameter: it increases from $\approx1.4$~mas (uncorrected) to $\approx1.6$~mas (corrected), moving towards the JSDC value of $\approx2.1$~mas. The same trend is found for HD\,139329, whose fitted diameter decreases from $\approx2.0$ to $\approx1.8$~mas against a catalogue value of $\approx1.0$~mas. The uncertainty on the fitted diameter is dominated by systematics rather than photon noise: the formal $1\sigma$ error from the covariance matrix of the merged fit is $0.023$--$0.032$~mas, but the scatter between the four BCD configurations reaches $0.016$--$0.046$~mas, so we adopt the quadratic sum of the two, $0.03$--$0.06$~mas, as our final error bar. In both cases the correction reduces the bias relative to the catalogue, although the absolute value is not fully recovered. The observing conditions were favourable over most of the sequence, with a mean seeing of $0.86\arcsec$ and a mean coherence time of $\tau_0 = 3.4$~ms, degrading to $\tau_0 \simeq 2.2$~ms only at the very end of the night. As GRA4MAT is expected to stabilise the transfer function down to $\tau_0 \simeq 2.5$~ms \cite{Lopez2022, Woillez2024}, most of our data lie in the regime where the transfer function should be insensitive to the atmosphere. This is confirmed epoch by epoch: HD\,139329, observed twice at $\tau_0 = 3.8$ and $2.8$~ms, shows a residual offset with respect to the JSDC diameter growing only from $+0.72$ to $+0.84$~mas, so that the $\tau_0$ degradation accounts for at most $\sim20\%$ of the offset. This epoch-to-epoch drift is nevertheless reduced by a factor of $2.5$--$4.5$ when the V-factor correction is applied ($0.33\rightarrow0.13$~mas for HD\,139329 and $0.22\rightarrow0.05$~mas for HD\,145250), showing that the V-factor does improve the stability of the transfer function against $\tau_0$ variations. The residual offset itself cannot be attributed to the atmosphere: a loss of coherence would bias all targets towards larger apparent diameters, whereas we measure offsets of opposite signs for the two calibrators ($+0.72$~mas for HD\,139329, $-0.53$~mas for HD\,145250). This anti-correlation is instead the expected signature of the mutual calibration of the two stars, each being used as the calibrator of the other. We note that our mutual calibration scheme, in which each calibrator is calibrated by the other, is a demonstrator of transfer-function stability rather than an absolute measurement of the diameters.

\begin{figure}[ht]
   \begin{center}
   \begin{tabular}{@{}c@{\hspace{2mm}}c@{}}
   \includegraphics[width=0.49\textwidth]{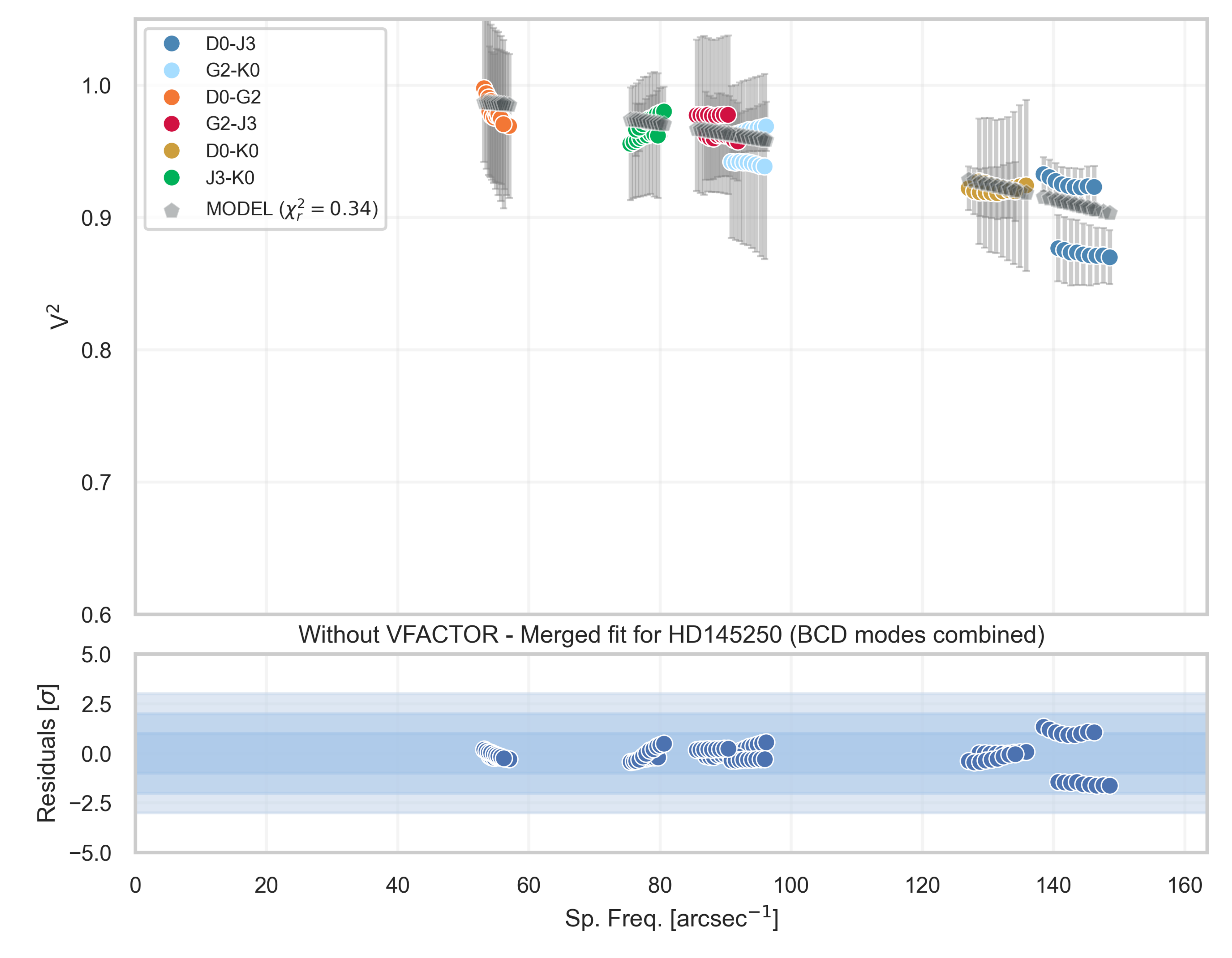} &
   \includegraphics[width=0.49\textwidth]{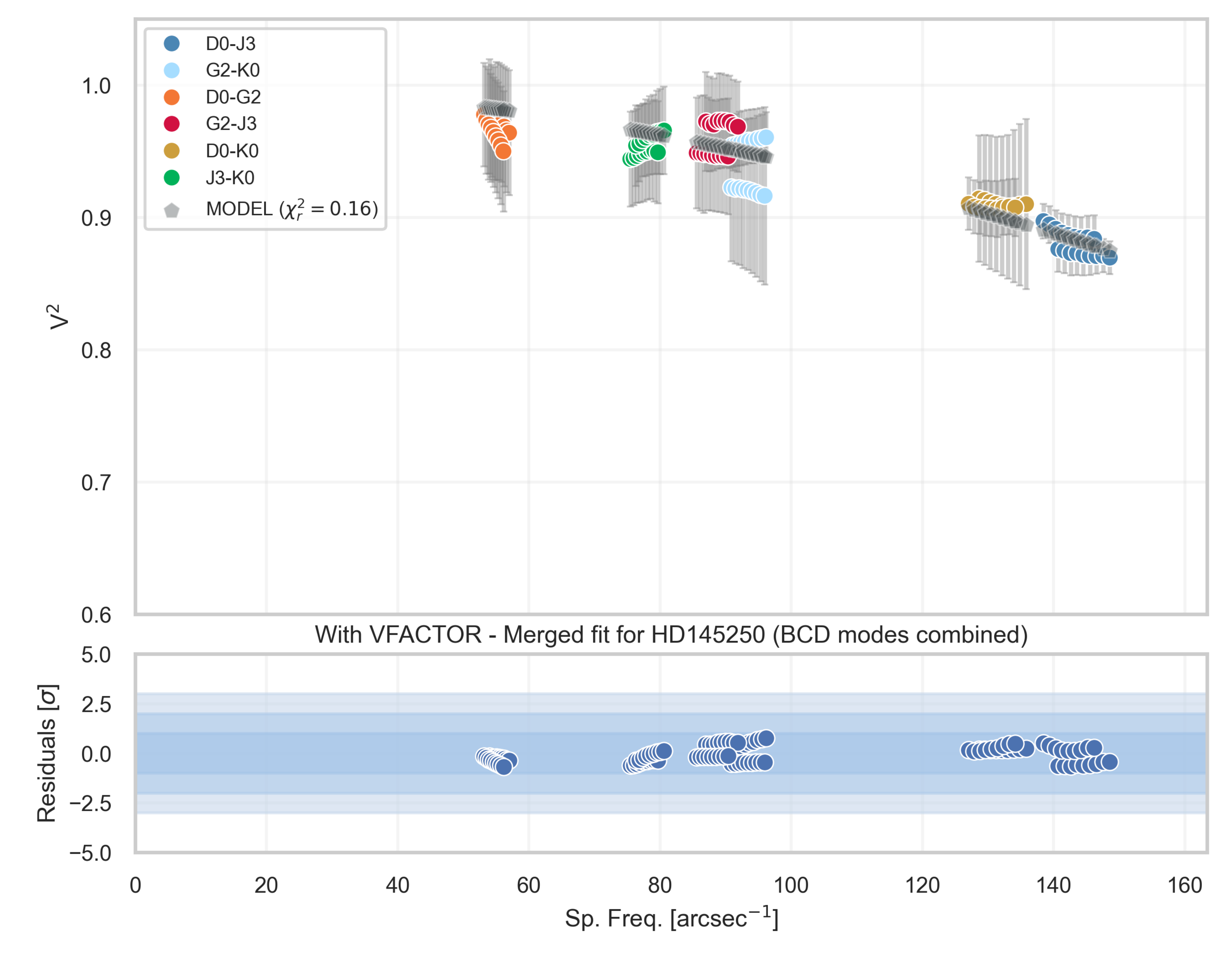} \\
   {\small (a) without V-factor} & {\small (b) with V-factor}
   \end{tabular}
   \end{center}
   \caption[v2fit]
   { \label{fig:v2fit}
Squared visibility $V^2$ versus spatial frequency for HD\,145250 (merged BCD modes),
before (a) and after (b) the V-factor correction, with the best-fitting uniform-disk
model (grey) and the per-point residuals in units of $\sigma$ (bottom sub-panels).
Both panels share the same vertical scale. After correction, a single model
reproduces all six baselines with $\chi^2_r = 0.16$.}
\end{figure}

The clearest improvement brought by the corrections is the internal consistency of the visibilities across baselines. Figure~\ref{fig:v2fit} shows the squared visibilities $V^2$ of the merged dataset of HD\,145250 as a function of spatial frequency, before and after correction, together with the best-fitting uniform-disk model. After correction, the six baselines are well described by a single model: the baseline-to-baseline dispersion is markedly reduced and the amplitude of the residuals decreases by a factor of $\sim1.4$ ($\chi^2_r = 0.16$, against $0.32$ before correction), with all points lying well within $1\sigma$ of the model over the full spatial-frequency range. The effect is most pronounced on the longest baseline, where the discrepancy present in the uncorrected data is fully removed. We note that both values of $\chi^2_r$ are below unity, indicating that the formal $V^2$ uncertainties are overestimated; we therefore use $\chi^2_r$ here as a relative measure of the residual dispersion rather than as an absolute goodness-of-fit indicator.

The benefit is most visible on faint targets observed under degraded conditions. For a representative L-band observation of a faint science target with the calibrator HD\,89998 (seeing $0.7''$, coherence time $\tau_0 = 9.8$~ms), enabling the V-factor correction recovers $+4.4\%$ of visibility, improves the L-band signal-to-noise ratio by $+18\%$, and reduces the statistical uncertainties by $-19\%$. These are precisely the regimes---low flux, marginal atmospheric conditions---where MATISSE measurements are otherwise least reliable, and where automated, consistent correction has the largest scientific impact.

\newpage
\section{CONCLUSION AND OUTLOOK}
\label{sec:conclusion}

MATISSE-py is already used for routine reductions of MATISSE data. By wrapping the official ESO \texttt{esorex} recipes behind a single, modular \texttt{matisse} command, it removes the practical barriers---manual SOF files, version-specific flags, sequential processing---that previously limited efficient reduction to expert users, while preserving the validated numerical core of the pipeline. A whole night can be reduced, calibrated, and inspected reproducibly in a small number of commands.

Crucially, the package does more than automate: it delivers measurable gains in data quality. The V-factor and P-factor corrections yield calibrator diameters that are consistent across BCD modes and recover signal that is otherwise lost on faint targets and under poor conditions. By consolidating the consortium's expertise into a tested, maintainable, and openly distributed package, MATISSE-py paves the way toward a modernized reduction ecosystem for MATISSE, supporting both future instrument commissioning and the broader user community. The software, its documentation, and a tutorial are publicly available,\cite{Soulain2026} and the package can be installed directly from PyPI.

\acknowledgments
This work was supported by the French National Research Agency (ANR) under the AGN-MELBa program, grant ANR-21-CE31-0011. This work was supported by the Action Spécifique Haute Résolution Angulaire (ASHRA) of CNRS/INSU co-funded by CNES. This work was supported by the Observatoire de la Côte d'Azur (OCA). ESO support in synergy with the MATISSE-UT-VLTI upgrades (Agreement No. 119258/ESO/26/129752/CBL). It is based on observations made with the VLTI/MATISSE instrument at the European Southern Observatory. The authors thank the MATISSE consortium for many years of accumulated operational expertise that this software aims to preserve.

\bibliography{report} 
\bibliographystyle{spiebib} 

\end{document}